\documentclass[conference, 9pt]{IEEEtran}
\usepackage{graphicx}

\usepackage{comment}  % Comments out with \begin{comment} ... \end{comment}

\usepackage{color}
\usepackage{url}
\usepackage[hidelinks]{hyperref}

\newcommand{\TODO}[1]{{\color{blue}\textit{#1}}}

\begin{document}
\title{Transparent Checkpointing for OpenGL Applications on GPUs}

%%% AUTHOR ORDERING:
%%%   David, Jun, Yue, Younes and Twinkle

\author{\IEEEauthorblockN{David Hou}
\IEEEauthorblockA{\textit{MemVerge, Inc.}\\
Milpitas, USA \\
david.hou@memverge.com}
\and
\IEEEauthorblockN{Jun Gan}
\IEEEauthorblockA{\textit{MemVerge, Inc.}\\
Milpitas, USA \\
jun.gan@memverge.com}
\and
\IEEEauthorblockN{Yue Li}
\IEEEauthorblockA{\textit{MemVerge, Inc.}\\
Milpitas, USA \\
yue.li@memverge.com}
\and
\IEEEauthorblockN{Younes El Idrissi Yazami}
\IEEEauthorblockA{\textit{Northeastern University}\\
Boston, USA \\
elidrissiyazami.y@northeastern.edu}
\and
\IEEEauthorblockN{Twinkle Jain}
\IEEEauthorblockA{\textit{Northeastern University}\\
Boston, USA \\
jain.t@northeastern.edu}
}

\maketitle
\thispagestyle{plain}
\pagestyle{plain}

\IEEEpeerreviewmaketitle

\begin{abstract}

This work presents transparent checkpointing of OpenGL applications, refining
the split-process technique~\cite{garg2019mana} for application in
GPU-based 3D graphics.  The split-process technique was earlier
applied to checkpointing MPI and CUDA programs, enabling
reinitialization of driver libraries.
%% It is planned to also use log-and-prune, as described
%% in an earlier approach to checkpointing of OpenGL by Kazemi \hbox{et
%% al.} toward checkpointing OpenGL. 
The presented design targets practical, checkpoint-package agnostic
checkpointing of OpenGL applications.
An early prototype is demonstrated on Autodesk Maya.
Maya is a complex proprietary media-creation software suite used with
large-scale rendering hardware for CGI (Computer-Generated Animation).
Transparent checkpointing of Maya provides critically-needed fault
tolerance, since
Maya is prone to crash when artists use some of its bleeding-edge
components.  Artists then lose hours of work in re-creating their
complex environment.
%% (something about versatility?)
\end{abstract}


\vspace{-0.2cm}
\section{Introduction}
\label{sec:introduction}

In complex media-creation programs such as Autodesk Maya~\cite{maya},
artists are faced with a crucial dilemma.  They can stay with the core
software and plugins, which are relatively robust.  But eventually,
for superior work, they are forced to use some third-party plugins
that can be prone to crash during normal operation.

A crash causes artists to lose hours of work and forces them to wait
multiple minutes to reload their project and continue working. These
interruptions put artists out of their flow at the most inopportune
times, and result in significant loss of productivity. Transparent
checkpoint/restart (C/R) technology is a promising tool for dealing
with these occurrences: by taking periodic snapshots of the editor
program in the background, the artist will be able to restore from a
recent checkpoint and quickly resume working.

Many media creation programs, including Maya, make extensive use of
GPUs to render 3D graphics using OpenGL~\cite{opengl} and to perform
heavier computations using CUDA and OpenCL. 
This typically occurs in the context of rendering farms (analogous to
traditional HPC clusters) for CGI (Computer-Generated Animation).
We will be focusing on
OpenGL in this paper, but the techniques described are reasonably
general and can be applied to any subsystem of this form. These APIs
are usually implemented by a vendor-specific library, which talks to
the hardware through various means.

The OpenGL API is structured as a state machine. The consumer (of the API) can
allocate and load various resources, such as shaders and textures, and operate
on the resources they have created. In handling these operations, the OpenGL
drivers will load some state into the GPU device. We must maintain this state
across checkpoint and restart. Unfortunately, current driver implementations do
not provide a convenient way to do this directly. To make matters worse,
driver-device communications are often closed-source, opaque, and unstable.

Luckily, the OpenGL API is well-defined and provides a clean,
deterministic interface. We can leverage this to capture a program's
entire OpenGL state by logging all OpenGL calls made by the program.
This provides a promising idea to support C/R of OpenGL programs:

\begin{itemize}
\item \textit{While the user program is running}: intercept and log all OpenGL
API calls to encapsulate the state of the OpenGL state machine.

\item \textit{When checkpointing}: drop all resources (VMAs, FDs, etc) related
to OpenGL drivers.

\item \textit{On restore}: recreate OpenGL drivers, and replay the logs to
restore driver state.
\end{itemize}

We are able to maintain a representation of the OpenGL drivers' state and reset
the drivers at will. One could say that we are checkpointing the OpenGL drivers'
state separately from the rest of the program.

We present two implementations of our system, one based on CRIU and
the other based on DMTCP. These two implementations share all core
functionalities, demonstrating the checkpoint-package agnostic nature
of our solution. These two implementations differ in their interaction
with their respective checkpointing packages in order to take
advantage of their different architectural properties.

There are two major problems to consider in order to enable
checkpoint-restart an OpenGL application: a)~reinitializing a fresh
OpenGL library on restart from a checkpoint image
(Section~\ref{sec:splitProcesses}); and b)~restoring the earlier state
of that reinitialized OpenGL library (Section~\ref{sec:log-replay}).
Following that discussion, an experimental evaluation for the Autodesk
Maya suite (Section~\ref{sec:experiment}) is presented.
Sections~\ref{sec:log-replay} and~\ref{sec:experiment} are based
primarily on the implementation using CRIU and VNC/VirtualGL.
Section~\ref{sec:dmtcp-experiment} presents a second implementation
using DMTCP with native handling of X instead of VNC, but only for the
GLX demo glxgears.  Lastly, related work
(Section~\ref{sec:relatedWork}) is presented.

% break into sections?
% introduce crac
% changes
% justifications
\section{Split processes for OpenGL and its kernel drivers}
\label{sec:resetting}
\label{sec:splitProcesses}

Naively, one would like to simply save the user-space memory of the
OpenGL library and restore it on restart.  This cannot work, since
on restart, the kernel drivers will be in a state inconsistent with
that of OpenGL.  And for natural reasons, there is no library call
in OpenGL to re-initialize the library and kernel drivers to a fresh
state.  So, the preferred solution is to load a fresh copy of OpenGL
during restart.  Any constructor functions in OpenGL will use knowledge
of internals to reset the kernel drivers at that time.

The approach taken is that of a \emph{split process}.
Jain \hbox{et al.}~\cite{jain2020crac} refined the split process
concept of MANA for MPI~\cite{garg2019mana} to apply to GPUs and CUDA.
That package, CRAC for CUDA~\cite{jain2020crac}, splits the memory of a
process into two regions:  application code and system libraries (e.g.,
network, MPI, CUDA, etc.).  All memory regions are tagged as upper half
(application code) or lower half (system libraries).  At the time of
checkpoint, only the upper half memory regions are saved.  At the time
of restart, a trivial lower-half application (with system libraries)
is launched, and the trivial application then restores the upper half memory
that was saved earlier.

The split-process approach
performs better than the well-known use of proxy or helper processes
(e.g., see
Kazemi \hbox{et al.}~\cite{kazemi2013transparent,kazemi2014v2transparent}).
In the proxy
approach, any OpenGL calls that use pointers require copying
to the proxy process the buffer referenced by the pointer.
With split processes,
a pointer is passed directly between upper and lower half, since they
share the same memory space.  Further efficiencies apply when managing
OpenGL resource ids.

While the OpenGL library is thread-safe, managing threads is non-trivial.
If the upper-half application has two threads, then there must
be two corresponding threads in the lower half.  Any use of thread-local
variables by the OpenGL library will be sensitive to this.

\begin{comment}
The GLX and X components of the system are less
well-behaved, sometimes returning malloc'd pointers, etc. which we need
to track and handle. These issues can be handled on a case-by-case
basis (see ``Virtualization of OpenGL graphics ids'' from
Section~\ref{sec:log-replay}).
\end{comment}

%% When we need to recreate the upper half, we can use the same linker-creation
%% mechanism from before. In fact, we can create as many pseudo-processes as we
%% want in the same address space, since there is nothing special about how they
%% obtain system resources.

%%%% I'm commenting this out for space reasons.  Not as important
%%%% as the other reasons for choosing this approach.  - Gene
%% We choose this split-process approach over using a separate process ...
%% 2) We only need to handle the behavior of the OpenGL drivers:
%% the methods (LD\_PRELOAD interception, patching) available to us to track
%% resources are not perfect (they require us to know where the user program is
%% calling into the kernel). Since we do not have a general way to deal with all
%% possible user applications, we want to mess with the user application as
%% little as possible.

Finally, the split process approach is C/R-package agnostic.
All OpenGL resource creation and deletion operations
by the driver-half libraries are captured and tracked independently
of the C/R package.  Further, the management of upper-/lower-halves
is independent of the C/R package.
To demonstrate this, the work has been implemented
twice (using the CRIU~\cite{criu} and DMTCP~\cite{ansel2009dmtcp} C/R
packages).

\begin{comment}
Two variants of the approach
have been implemented in this work, within the framework of
CRIU~\cite{criu} and DMTCP~\cite{ansel2009dmtcp}, to show that the
approach is general. Large parts of the code are the same in the two
variants.

The approach further opens up interesting use cases in server
virtualization: We can unload and recreate the drivers without the
help of a checkpointing package, and allows recovery of OpenGL states
on different machines with different OpenGL implementations and
different types of GPUs.
\end{comment}

\section{Log-replay to restore OpenGL state}
\label{sec:log-replay}

%% GLX API to manage the relationships between OpenGL rendering contexts
%% and application windows. GLX is an extension to the X-Windows protocol,
%% and it can take advantage of that protocol's inherent remote display
%% capabilities. In this mode of operation, referred to as "indirect
%% rendering", the OpenGL commands are encapsulated inside of the X-Windows
%% protocol stream and sent to an X-Windows server running on a remote
%% machine
Section~\ref{sec:resetting} described how to reinitialize the OpenGL
library and kernel drivers on restart.  Broadly, there are two
important libraries, OpenGL~\cite{shreiner2013opengl} and
GLX~\cite{humphreys1999distributed}, which are responsible,
respectively, for: (i)~state machines for rendering; and (ii)~managing
their interaction with X-Windows.  We are interested in providing the
user program this interface in a way that is consistent across
unload/reload operations.

To restore OpenGL state correctly during log-replay, we implement
virtualization of graphics IDs. The graphic IDs are not assigned
deterministically, and can change between checkpoint and restart. For
example, \texttt{glCreateShader($\ldots$)} returns a value with type
\texttt{GLuint}, which is an ID pointing to a graphic shader object.
We may log the ID saved in the user code, but on replay, we may
receive a different ID.  The solution is to maintain a translation
table between a virtual ID and the real ID returned by the current
OpenGL library.  The virtual ID is saved in the user code.  Any OpenGL
call using a virtual ID is automatically translated to the real ID of
the current OpenGL. On restart, the virtual-to-real table is updated
to use the real ID returned by the newly loaded OpenGL library.

GLX is primarily responsible for setting up an X11 Window that is
responsive to X11 events. The OpenGL library creates the graphics
image within a frame buffer provided by X11. We are able to save the
entire state of these libraries by a log-and-replay system for
capturing calls made to the two libraries. In addition, we need to
handle the connection to the X server, since user program will hold
coordinated state (e.g., created windows) with the X server. The
standard method for dealing with this for non-OpenGL programs is to
checkpoint the X server along with the user program, so that both the
client and server state are captured together~\cite{criu,
  ansel2009dmtcp}. However, these X servers generally do not support
the GLX extension, so we cannot normally use GPU-backed OpenGL
rendering with them. Luckily, there exists an off-the-shelf solution
to this, VirtualGL. VirtualGL separates rendering from the X server
implementation one by using an alternative X server to do the
rendering and passing the resultant bitmap images to the desired X
server. Therefore, we just need to shut down the alternative X server
connection before checkpoint and restore it on restart along with the
OpenGL drivers.

\section{Experimental Evaluation}
\label{sec:experiment}

% todo: talk about opengl coverage? --- opengl xml spec
% todo: maybe talk about xml spec -> x log/replay too?
% todo: problems with our setup?

% put picture
% 3 tests, data comparison
% environment

% runtime overhead in progress

Our prototype has been applied to Autodesk Maya 2020, a complex
computer graphics design software widely deployed in the movie and
entertainment industry. One of our goals is to demonstrate the fast
restart capability that our approach brings to Maya for better
crash-recovery. All the tests ran on a 2-socket compute server running
CentOS~$7.6$ with Intel Xeon Gold~$5220$ CPU ($2.2$~GHz, $18$~cores/socket),
$192$~GB of RAM, two $1.6$~TB NVMe SSDs and one Nvidia
RTX $4000$ graphics card. We use CRIU as the C/R package, a VNC session to
support checkpointing of the X windows, and VirtualGL is used to
support GPU-accelerated OpenGL with VNC.

We use Maya to first load a
moderately-sized model from the Disney's Moana Island Scene
dataset~\cite{moana}.  We measure the loading time for both baseline
(without our system), and using log-replay for OpenGL calls (with our
system).
In our system, we checkpoint (using the CRIU variant of our
software), kill the Maya process, and restart.
We show that our checkpoint-restart time is actually \emph{shorter}
than the time for the baseline to:
relaunch Maya, wait for the initialization to
finish and reload the same model from storage.

\begin{figure}[htbp]
  \centering
    \includegraphics[width=\linewidth]{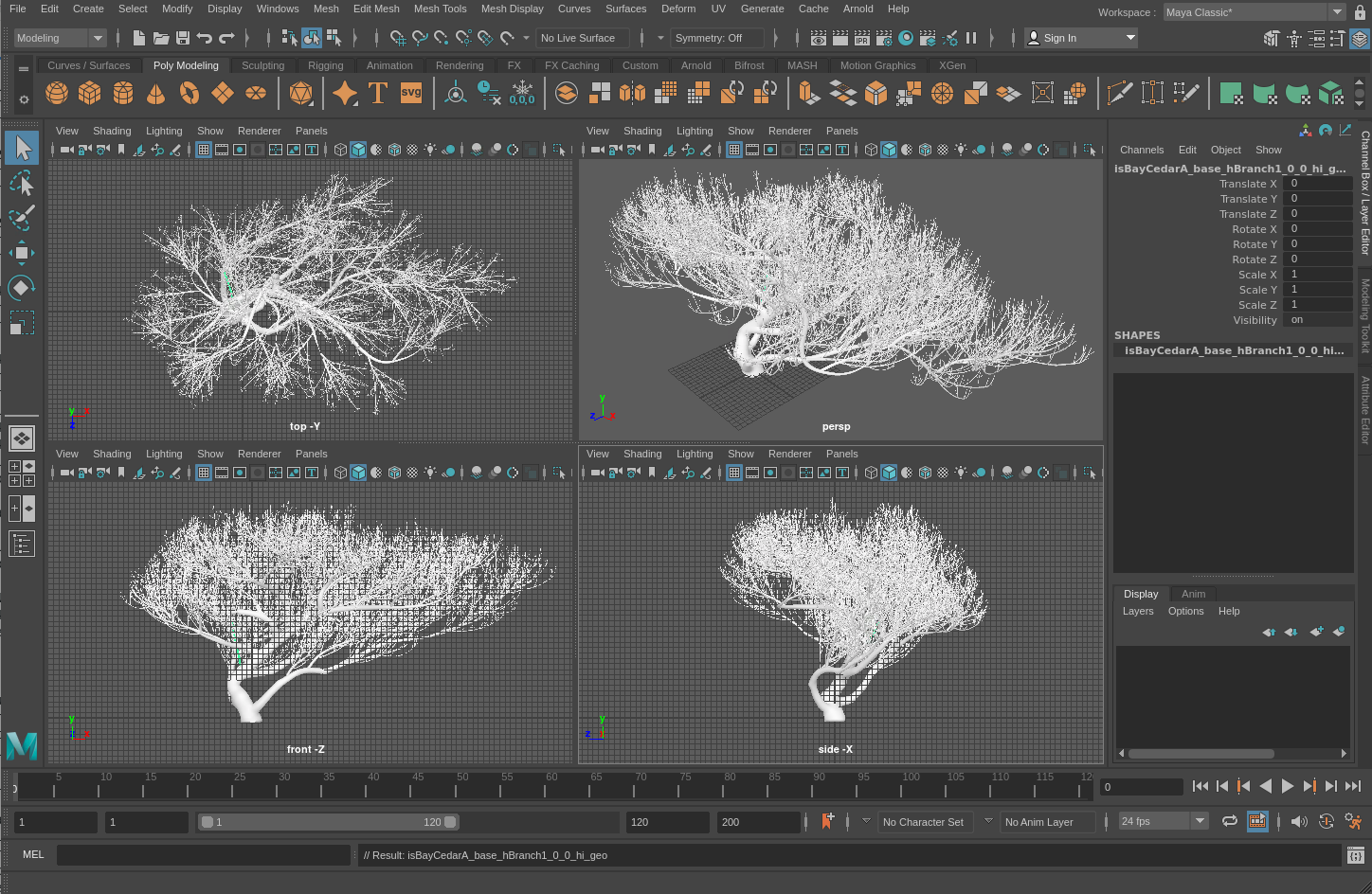}
  \caption{\label{fig:maya} Maya after loading a model from
    the Moana Island Scene dataset.}
\end{figure}

Figure~\ref{fig:maya} shows Autodesk Maya loaded with the model
\texttt{isBayCedarA1.obj} ($121$~MB).
 
\begin{figure}[htbp]
  \centering
    \includegraphics[width=\linewidth]{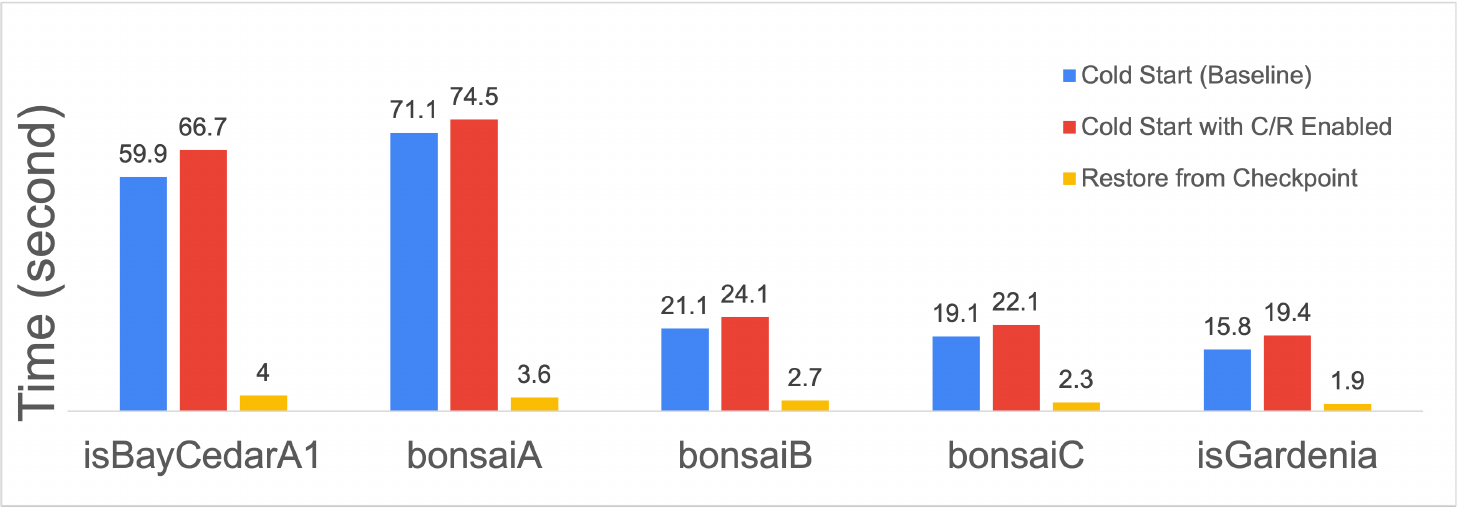}
  \caption{\label{fig:results} Maya baseline cold start time, cold
    start time with C/R enabled, and checkpoint-restart time on
    different models.}
\end{figure}

For the baseline, launching Maya and this model takes $60$~s. We are able to
restart Maya at this state from a checkpoint image on disk in just $4$~s.
Figure~\ref{fig:results} shows baseline vs\hbox{.} C/R for various models from
the dataset, and shows a clear advantage of our approach on restart speed.
 
%% Can we also check in the source from which the graph was created? - Gene

The library logging, use of VNC and VirtualGL~\cite{vgl} during the
normal operation of the OpenGL program necessarily incur performance
overhead. Preliminary results show up to $10\%$ increase on cold start
time (Figure~\ref{fig:results}) and a noticeable viewport
frames-per-second (FPS) penalty when interacting with the
model. However, we are still able to achieve a very usable FPS
on models that were tested. We plan to improve upon this in
the future by reducing the number of extra transport layers
introduced by the facilities above, and by introducing log
pruning~\cite{kazemi2013transparent}.

\section{Second Implementation using DMTCP}
\label{sec:dmtcp-experiment}

As explained in the introduction, two implementations were created
using two different checkpointing packages:  CRIU and DMTCP.
Both implementations use the same split-process approach described in
Section~\ref{sec:splitProcesses}.  This shows the generality of the
approach.

The DMTCP design does not require VNC or VirtualGL for its
operation.  However, the DMTCP implementation is not as far advanced,
and so it is demonstrated for the well-known GLX demo glxgears, instead
of for Maya. The two implementations share code for the log-and-replay
that was described in Section~\ref{sec:log-replay}.

\begin{figure}[htbp]
  \centering
    \includegraphics[width=\linewidth]{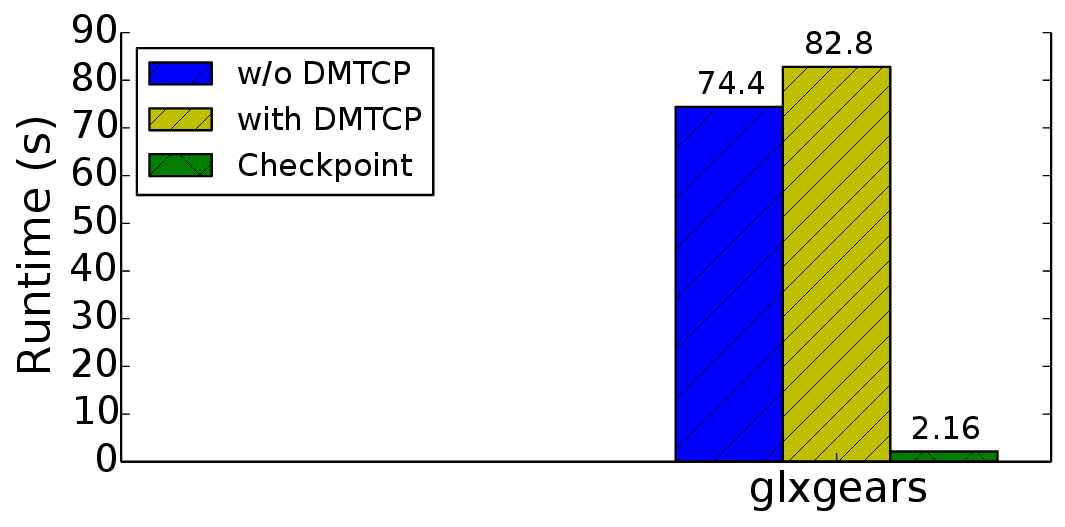}
  \caption{\label{fig:glxgears}glxgears runtime overhead (with and
  w/o DMTCP)
	and checkpoint time.}
\end{figure} 

Glxgears normally runs infinitely and prints back out the frames
per-second (FPS) information. However, we modified glxgears to loop
for $10,000$ times for these experiments.  The experiments were run
inside a virtual machine using version 3.0 of the mesa OpenGL
implementation. Mesa emulates the graphics in software without the
use of the GPU.

Glxgears was run without DMTCP to establish a baseline (w/o DMTCP),
and then under DMTCP's control (with DMTCP). In Figure~\ref{fig:glxgears}
we can see that glxgears incurred an overhead of 8\% when running
under DMTCP's control. This is similar to the Maya experiments and is
due to the many OpenGL calls made by glxgears.

The 8\% overhead should be either due to the startup time of DMTCP
or due to switching between the application and driver's half.  A
quick experiment of doubling the glxgears loop's number of iterations
showed that the overhead remained at 8\%, showing that DMTCP startup
time was insignificant.

Hence, switching back and forth between application and driver's half is
responsible for the 8\% overhead.  When glxgears makes an OpenGL
call, the call is redirected from the application half to the
driver's half. The 8\% overhead can be improved (lowered) by using
Linux's upcoming FSGSBASE as described in~\cite{jain2020crac}.

\begin{comment}
( checkpoint time should be checked against
the restart time) currently the restart time takes as much as the
app runtime till checkpoint. If app1 run 1min and we take a
checkpoint, restart time will be 1 min (since we are replaying all
the calls)
\end{comment}

\section{Related Work} \label{sec:relatedWork} \label{sec:x-windows}

\begin{comment}
\TODO{If there's no room for this ``related work'' section, we can
      comment it out, and leave it for the full paper.}
\end{comment}

There is a surprisingly long history of checkpointing OpenGL (3D~graphics)
for X-Windows.
In 2007, Lagar-Cavilla \hbox{et al.} presented VMGL~\cite{lagar2007vmm},
demonstrating vendor-independent checkpoint-restart for OpenGL
version~1.5.  This landmark result employed a shadow device driver
to model the OpenGL state and restore it on restart.  This work was
VMM-independent (independent of the virtual machine).
In 2010, Lin \hbox{et al.}~\cite{lin2010opengl} showed live migration
of GPU-based 3D graphics between machines.
In 2013, Kazemi
\hbox{et al.}~\cite{kazemi2013transparent,kazemi2014v2transparent} showed
checkpoint-restart for OpenGL version~3, using a record-prune-replay
technique.  In future work, we intend to integrate record-prune-replay.

\section*{Acknowledgment}
This work was partially supported by National Science Foundation Grant
OAC-1740218 and a grant from Intel Corporation.

% todo: bibliography
%\vspace{-0.1cm}
%\bibliographystyle{IEEEtranS}
%\bibliography{allbib}

%\begin{thebibliography}{1}
%\end{thebibliography}
%\IEEEtriggeratref{8}  % split columns starting at this citation number.
\bibliographystyle{IEEEtran}
\bibliography{ckpt-opengl}
\end{document}